\documentclass[10pt,conference]{IEEEtran}
\usepackage[cmex10]{amsmath}
\usepackage{amssymb,amsthm}
\usepackage{url}
\usepackage[dvips]{graphicx}
\usepackage{psfrag}
\usepackage{stfloats}

\newtheorem{theorem}{Theorem}
\newtheorem{lemma}{Lemma}

\def\Pb{\mathrm{P_b}}
\def\Pn{\mathbb{P}_n}

\title{Finite-Length Analysis of Irregular Expurgated LDPC Codes under Finite Number of Iterations}

\author{\IEEEauthorblockN{Ryuhei Mori\IEEEauthorrefmark{1},
Toshiyuki Tanaka\IEEEauthorrefmark{1},
Kenta Kasai\IEEEauthorrefmark{2}, and
Kohichi Sakaniwa\IEEEauthorrefmark{2}}
\IEEEauthorblockA{\IEEEauthorrefmark{1}Department of Systems Science\\
Kyoto University,
Kyoto 606-8501, Japan\\
Email: rmori@sys.i.kyoto-u.ac.jp,~tt@i.kyoto-u.ac.jp}
\IEEEauthorblockA{\IEEEauthorrefmark{2}Department of Communications and Integrated Systems\\
Tokyo Institute of Technology,
Tokyo 152-8552, Japan\\ Email: \{kenta, sakaniwa\}@comm.ss.titech.ac.jp}
}

\begin{document}
\maketitle

\begin{abstract}
Communication over the binary erasure channel (BEC) using low-density parity-check (LDPC) codes and belief propagation (BP) decoding is considered.
The average bit error probability of an irregular LDPC code ensemble after a fixed number of iterations converges to a limit, 
which is calculated via density evolution, as the blocklength $n$ tends to infinity.
The difference between the bit error probability with blocklength $n$ and the large-blocklength limit behaves asymptotically like $\alpha/n$,
where the coefficient $\alpha$ depends on the ensemble, the number of iterations and the erasure probability of the BEC\null. 
In~\cite{4595026}, $\alpha$ is calculated for regular ensembles.
In this paper, $\alpha$ for irregular expurgated ensembles is derived.
It is demonstrated that convergence of numerical estimates of $\alpha$ to the analytic result is significantly fast for irregular unexpurgated ensembles.
\end{abstract}

\setcounter{equation}{0}
\section{Introduction}
In this paper, we consider communication over the binary erasure channel (BEC) using low-density parity-check (LDPC) codes and belief propagation (BP) decoding.
It is important to predict the average bit error probability of an LDPC code ensemble for designing a practical code.
The bit error probability of an ensemble is determined by a blocklength, erasure probability of a channel, and the number of BP iterations.
Let $\Pb(n,\epsilon,t)$ denote the bit error probability of an ensemble of codes of blocklength $n$ over the $\text{BEC}(\epsilon)$ after $t$ BP iterations.
The limit of large blocklength of the bit error probability, denoted by $\Pb(\infty,\epsilon,t)$, is obtained easily via \textit{density evolution}~\cite{910577}.
An important consequence is that there exists a threshold erasure probability of the channel, $\epsilon_\text{BP}$,
such that $\lim_{t\to\infty}\Pb(\infty,\epsilon,t)=0$ if $\epsilon<\epsilon_\text{BP}$, and
$\lim_{t\to\infty}\Pb(\infty,\epsilon,t)>0$ if $\epsilon>\epsilon_\text{BP}$.

Although the analysis in the large-blocklength limit is easy,
estimation of performance for finite blocklengths is much more complicated~\cite{1003839}~\cite{1023275}~\cite{RiU02/LTHC}. 
These analyses require high computational costs which grow like a power of the blocklength and like an exponential of the number of degrees.

A large-$n$ asymptotic analysis is useful for avoiding high computational complexity.
An asymptotic analysis of the large-$t$ limit of the bit error probability below the threshold is shown in~\cite{1715529}.
However, the large-$n$ asymptotic analysis in the limit $t\to\infty$ breaks down near the threshold:
If the large-blocklength limit of the bit error probability is discontinuous at the threshold,
the convergence is not uniform in regions including the threshold.
If not, the coefficient of $1/n$ diverges as $\epsilon$ approaches the threshold from below.  
The \textit{scaling-law}-based method~\cite{Am07} has been shown to be useful 
near the threshold.

The large-$n$ asymptotic analysis with finite $t$~\cite{4595026} 
provides an alternative useful approach.  
Indeed, for a finite $t$, 
the convergence of $\Pb(n,\epsilon,t)$ to the limit $\Pb(\infty,\epsilon,t)$ 
as $n\to\infty$ 
is uniform on $\epsilon\in[0,1]$.
Hence, the asymptotic expansion 
\begin{equation}
\Pb(n,\epsilon,t) = \Pb(\infty,\epsilon,t) + \alpha(\epsilon,t)\frac1{n} + o\left(\frac1{n}\right)\label{apalpha}
\end{equation}
is well behaved, so that 
approximations using~\eqref{apalpha} while ignoring the $o(n^{-1})$ term should be accurate for all $\epsilon$ uniformly if the blocklength is sufficiently large.
The coefficient $\alpha(\epsilon,t)$ of $n^{-1}$ for regular ensembles is obtained in~\cite{4595026}.
In this paper, $\alpha(\epsilon,t)$ for irregular expurgated ensembles is derived.
In Section~\ref{sec:prev}, outline of calculation of $\alpha(\epsilon,t)$ is described.
In Section~\ref{sec:betai}, a generalization to irregular ensembles is shown.
In Section~\ref{sec:gexp}, a further generalization to irregular expurgated ensembles is outlined.
In Section~\ref{calsim}, numerical calculation results of $\alpha(\epsilon,t)$ for irregular expurgated ensembles
and simulation results corresponding to $\alpha(\epsilon,t)$ for irregular unexpurgated ensembles are shown.
Finally,  we conclude this paper in Section~\ref{cncld}.

\section{Outline of calculation of $\alpha(\epsilon,t)$}\label{sec:prev}
In this paper, we consider the standard irregular LDPC code ensemble $(\lambda(x),\rho(x))$
where the degree distributions of variable and check nodes are fixed and
where the set of edges is chosen uniformly from all possible choices.
An error occurrence after $t$ iterations depends only on a neighborhood graph $G$ of depth $t$ and
channel outputs at variable nodes in $G$.
Hence, the bit error probability of an irregular ensemble is calculated as
\begin{equation*}
\Pb(n,\epsilon,t) = \sum_{G\in\mathcal{G}_t} \Pn(G)\Pb(\epsilon,G)
\end{equation*}
where $\mathcal{G}_t$ denotes the set of all neighborhood graphs of depth $t$,
where $\Pn(G)$ denotes the probability that the neighborhood graph $G$ is generated in the code ensemble considered,
and where $\Pb(\epsilon,G)$ denotes the error probability of the root node of $G$ after $t$ iterations
when each message into variable node in $G$ is transmitted over the $\text{BEC}(\epsilon)$.
%
If we distinguished all sockets, the probability that a neighborhood graph $G$ is generated would be
\begin{equation*}
\frac1{nE(E-1)\dotsm (E-(k-1))}
\end{equation*}
where $E$ and $k$ are the numbers of edges in the whole Tanner graph and in the neighborhood graph $G$, respectively.
This distinction is finer than necessary for our purpose, 
since we do not have to distinguish nodes of the same degrees, 
so that the following marginalized probability is considered.
We order sockets in the same node and number them, 
and then identify the nodes with the same degree.
We further identify sockets by their number in the cyclic sense.
On this identification, the marginalized probability of particular neighborhood graph $G$ is
\begin{equation}
\Pn(G) := nL_{u}\frac{\prod_i\prod_{l=0}^{v_i-1}(E\lambda_i-l)\prod_j\prod_{l=0}^{c_j-1}(E\rho_j - l)}{nE(E-1)\dotsm (E-(k-1))}
\label{eq:pnb}
\end{equation}
where $u$ denotes the degree of the root node and
where $v_i$ and $c_i$ denote the numbers of variable and check nodes of degree $i$ in the neighborhood graph, respectively.

The following lemma is an important consequence of \eqref{eq:pnb}.
\begin{lemma} \label{lemma:kcycles}
For a neighborhood graph $G$ which has $c$ cycles,
\begin{equation*}
\Pn(G) = \Theta(n^{-c})
\end{equation*}
as $n\to\infty$ while the number of iterations is fixed.
\end{lemma}
\noindent
From Lemma~\ref{lemma:kcycles}, it holds that
\begin{equation*}
\Pb(\infty,\epsilon,t)
= \sum_{G\in\mathcal{T}_t} \mathbb{P}_\infty(G)\Pb(\epsilon, G) \text{,}
\end{equation*}
where $\mathbb{P}_\infty(G):=\lim_{n\to\infty}\Pn(G)$ and 
where $\mathcal{T}_t$ denotes the set of all cycle-free neighborhood graphs of depth $t$.
From this fact, the limit of the bit error probability $\Pb(\infty,\epsilon,t) := \lim_{n\to\infty}\Pb(n,\epsilon,t)$ is calculated recursively.
\begin{lemma}[Density evolution~\cite{910577}]
Let $Q_\epsilon(t)$ denote erasure probability of messages into check nodes at $t$-th iteration and
$P_\epsilon(t)$ denote erasure probability of messages into variable nodes at $t$-th iteration in the limit $n\to\infty$.
Then
\begin{align*}
\Pb(\infty,\epsilon,t) &= \epsilon L(P_\epsilon(t))\text{,}\\
Q_\epsilon(t) &= \epsilon\lambda(P_\epsilon(t-1))\text{,}\\
P_\epsilon(t) &=
\begin{cases}
1,&\text{if } t = 0\\
1-\rho(1-Q_\epsilon(t)),&\text{otherwise.}
\end{cases}
\end{align*}
\end{lemma}

From Lemma~\ref{lemma:kcycles}, we can see that the second dominant term is $\Theta(n^{-1})$.
The coefficient of $n^{-1}$, defined as
\begin{equation*}
\alpha(\epsilon,t) := \lim_{n\to\infty} n(\Pb(n,\epsilon,t) - \Pb(\infty,\epsilon,t)) \text{,}
\end{equation*}
determines the speed of convergence of $\Pb(n,\epsilon,t)$ to $\Pb(\infty,\epsilon,t)$ as $n$ tends to infinity.
Furthermore, Lemma~\ref{lemma:kcycles} tells us that
$\alpha(\epsilon,t)$ can be decomposed into two components as follows:
\begin{align*}
\alpha(\epsilon,t) &= \lim_{n\to\infty} n\left(\sum_{G\in\mathcal{T}_t} \Pn(G)\Pb(\epsilon,G) - \Pb(\infty,\epsilon,t)\right)\nonumber\\
& \hspace{9em} +\lim_{n\to\infty} n\sum_{G\in\mathcal{S}_t} \Pn(G)\Pb(\epsilon,G)\nonumber\\
&=: \beta(\epsilon,t)+\gamma(\epsilon,t) \text{,}
\end{align*}
where $\mathcal{S}_t$ denotes the set of all single-cycle neighborhood graphs of depth $t$
and where $\beta(\epsilon,t)$ and $\gamma(\epsilon,t)$ represent contributions
from cycle-free and single-cycle neighborhood graphs, respectively.
In~\cite{4595026}, $\gamma(\epsilon,t)$ was derived for irregular ensembles,
whereas $\beta(\epsilon,t)$ was derived only for regular ensembles.
In this paper, $\beta(\epsilon,t)$ for irregular ensembles is shown.
Furthermore, we consider the expurgated ensembles defined in~\cite{1715529} and
outline derivation of $\alpha(\epsilon,t)$ for the irregular expurgated ensembles.

\section{$\beta(\epsilon,t)$ for irregular ensembles}\label{sec:betai}
The contribution $\beta(\epsilon,t)$ of cycle-free neighborhood graphs is calculated as
\begin{align*}
\beta(\epsilon,t)&:=\lim_{n\to\infty} n\left(\sum_{G\in\mathcal{T}_t} \Pn(G)\Pb(\epsilon,G) - \Pb(\infty,\epsilon,t)\right)\\
&=\sum_{G\in\mathcal{T}_t} \left[\lim_{n\to\infty} n\left(\Pn(G)-\mathbb{P}_\infty(G)\right)\right]\Pb(\epsilon,G)\text{.}
\end{align*}
The contribution of a neighborhood graph $G$ to $\beta(\epsilon,t)$ is obtained as
\begin{multline*}
\lim_{n\to\infty} L_{|u|}\prod_{v\in \mathcal{V}(G)\backslash u} \lambda_{|v|} \prod_{c\in \mathcal{C}(G)} \rho_{|c|} \Pb(\epsilon,G)\\
\times n\left(\frac{\prod_i \prod_{l=0}^{v_i-1}\left(E-l\frac{i}{\lambda_i}\right) \prod_j \prod_{l=0}^{c_j-1}\left(E-l\frac{j}{\rho_j}\right)}{\prod_{i=0}^{k-1}(E-i)}-1\right)\\
= L_{|u|}\prod_{v\in \mathcal{V}(G)\backslash u} \lambda_{|v|} \prod_{c\in \mathcal{C}(G)} \rho_{|c|} \Pb(\epsilon,G)\frac1{2L'(1)}\\
\times\left(k(k-1)-\sum_i\frac{i}{\lambda_i}v_i(v_i-1)-\sum_j\frac{j}{\rho_j}c_j(c_j-1) \right)
\text{,}
\end{multline*}
where $\mathcal{V}(G)$ denotes the set of variable nodes in $G$, where $\mathcal{C}(G)$ denotes the set of check nodes in $G$
and where $|m|$ denotes degree of node $m$.
Hence, $\beta(\epsilon,t)$ is obtained by taking expectation $\mathbb{E}_t[\cdot]$ on tree ensemble~\cite{RiU05/LTHC} of depth $t$ from node perspective
\begin{multline*}
\frac1{2L'(1)}\biggl[\mathbb{E}_t[K(K-1)P] - \sum_i\frac{i}{\lambda_i}\mathbb{E}_t[V_i(V_i-1)P]\\
- \sum_j\frac{j}{\rho_j}\mathbb{E}_t[C_j(C_j-1)P]\biggr]\text{,}
\end{multline*}
where $K$ denotes the number of edges in $G$, $V_i$ denotes the number of variable nodes of degree $i$, $C_j$ denotes the number of check nodes of degree $j$,
and $P$ denotes the erasure probability of the root node after $t$ BP iterations.
The three expectations are obtained using generating functions as
\begin{align*}
\mathbb{E}_t[K(K-1)P] &= \left.\frac{\partial^2 \mathbb{E}_t[x^KP]}{\partial x^2}\right|_{x=1}\text{,}\\
\mathbb{E}_t[V_i(V_i-1)P] &= \left.\frac{\partial^2 \mathbb{E}_t[x^{V_i}P]}{\partial x^2}\right|_{x=1}\text{,}\\
\mathbb{E}_t[C_j(C_j-1)P] &= \left.\frac{\partial^2 \mathbb{E}_t[x^{C_j}P]}{\partial x^2}\right|_{x=1}\text{.}
\end{align*}
These generating functions $\mathbb{E}_t[x^KP]$, $\mathbb{E}_t[x^{V_i}P]$ and $\mathbb{E}_t[x^{C_j}P]$ are obtained using the following lemma.

\begin{lemma}\label{gen}
\begin{equation*}
\mathbb{E}_t\left[\prod_ky_k^{V_k}\prod_lz_l^{C_l}P\right] = \epsilon \mathfrak{L}(F(t))\text{,}\\
\end{equation*}
where
\begin{align*}
F(t) &:= \begin{cases}
1,&\text{if } t = 0\\
f(t) - \mathcal{P}(G(t)),& \text{otherwise,}\\
\end{cases}\\
G(t) &:=
g(t) - \epsilon \mathcal{L}(F(t-1))\text{,}\\
f(t) &:= \begin{cases}
1,&\text{if }t=0\\
\mathcal{P}(g(t)),&\text{otherwise,}
\end{cases}\\
g(t) &:= \mathcal{L}(f(t-1))\text{,}
\end{align*}
and where
\begin{align*}
\mathfrak{L}(x) &:= \sum_i L_i y_i x^i \text{,}&
\mathcal{L}(x) &:= \sum_i \lambda_i y_i x^{i-1} \text{,}\\
\mathcal{P}(x) &:= \sum_j \rho_j z_j x^{j-1} \text{.}
\end{align*}
\end{lemma}
\noindent
Using this generating function, those three generating functions are obtained as
\begin{align*}
\mathbb{E}_t[x^KP] &= \frac1{x}\left.\mathbb{E}_t\left[\prod_ky_k^{V_k}\prod_lz_l^{C_l}P\right]\right|_{y_k=x,z_l=x\text{ for all } k,l}\text{,}\\
\mathbb{E}_t[x^{V_i}P] &= \left.\mathbb{E}_t\left[\prod_ky_k^{V_k}\prod_lz_l^{C_l}P\right]\right|_{y_i=x, y_k=1, z_l=1\text{ for all } k\ne i, l}\text{,}\\
\mathbb{E}_t[x^{C_j}P] &= \left.\mathbb{E}_t\left[\prod_ky_k^{V_k}\prod_lz_l^{C_l}P\right]\right|_{z_j=x, y_k=1, z_l=1\text{ for all } k, l\ne j}\text{.}
\end{align*}

From Lemma \ref{gen}, the derivatives of the generating functions can be computed recursively.
Since derivation is straightforward, we only show the results.

\setcounter{equation}{5}
\begin{theorem}\label{beta}
$\beta(\epsilon,t)$ for $(\lambda(x),\rho(x))$-irregular ensembles is calculated as
\begin{multline*}
\beta(\epsilon,t) = \frac1{2L'(1)}\biggl[\mathbb{E}_t[K(K-1)P]\\
-\sum_i\frac{i}{\lambda_i}\mathbb{E}_t[V_i(V_i-1)P]-\sum_j\frac{j}{\rho_j}\mathbb{E}_t[C_j(C_j-1)P]\biggr]\text{,}
\end{multline*}
where $\mathbb{E}_t[K(K-1)P]$, $\mathbb{E}_t[V_i(V_i-1)P]$ and $\mathbb{E}_t[C_j(C_j-1)P]$ are calculated by (\ref{bk}), (\ref{bv}) and (\ref{bc}), respectively.
\begin{align*}
f'(t) &:= \begin{cases}
0\text{,}&\text{if }t=0\\
1+\rho'(1)g'(t)\text{,}&\text{otherwise,}
\end{cases}\\
g'(t) &:= 1+\lambda'(1)f'(t-1)\text{,}\\
F'(t) &:=\begin{cases}
0\text{,}&\text{if } t = 0\\
f'(t) - \rho(1-Q_\epsilon(t))\\
\quad - \rho'(1-Q_\epsilon(t))G'(t)\text{,}&\text{otherwise,}\\
\end{cases}\\
G'(t) &:= g'(t) - \epsilon \lambda(P_\epsilon(t-1))\\
&\quad - \epsilon \lambda'(P_\epsilon(t-1))F'(t-1)\text{,}
\end{align*}
\begin{align*}
f''(t) &:= \begin{cases}
0\text{,}&\text{if }t=0\\
2\rho'(1)g'(t)\\
\quad +\rho''(1)g'(t)^2+\rho'(1)g''(t)\text{,}&\text{otherwise,}
\end{cases}\\
g''(t) &:= 2\lambda'(1)f'(t-1)+\lambda''(1)f'(t-1)^2\\
&\quad +\lambda'(1)f''(t-1)\text{,}\\
F''(t) &:= \begin{cases}
0\text{,}&\text{if } t = 0\\
f''(t) - 2\rho'(1-Q_\epsilon(t))G'(t)\\
\quad - \rho''(1-Q_\epsilon(t))G'(t)^2\\
\quad - \rho'(1-Q_\epsilon(t))G''(t)\text{,}&\text{otherwise,}\\
\end{cases}\\
G''(t) &:=
g''(t) - 2\epsilon\lambda'(P_\epsilon(t-1))F'(t-1)\\
&\quad - \epsilon\lambda''(P_\epsilon(t-1))F'(t-1)^2\\
&\quad - \epsilon\lambda'(P_\epsilon(t-1))F''(t-1)\text{,}
\end{align*}
\begin{equation}
\mathbb{E}_t[K(K-1)P] = \epsilon L''(P_\epsilon(t))F'(t)^2 + \epsilon L'(P_\epsilon(t))F''(t)\text{,}\label{bk}
\end{equation}
\begin{align*}
f_v'(t,i) &:= \begin{cases}
0\text{,}&\text{if }t=0\\
\rho'(1)g_v'(t,i)\text{,}&\text{otherwise,}
\end{cases}\\
g_v'(t,i) &:= \lambda'(1)f_v'(t-1,i) + \lambda_i\text{,}\\
F_v'(t,i) &:=\begin{cases}
0\text{,}&\text{if } t = 0\\
f_v'(t,i) - \rho'(1-Q_\epsilon(t))G_v'(t,i)\text{,}&\text{otherwise,}\\
\end{cases}\\
G_v'(t,i) &:= g_v'(t,i) - \epsilon \lambda'(P_\epsilon(t-1))F_v'(t-1,i)\\
&\quad - \epsilon \lambda_i P_\epsilon(t-1)^{i-1}\text{,}
\end{align*}

\begin{align*}
f_v''(t,i) &:= \begin{cases}
0\text{,}&\text{if }t=0\\
\rho''(1)g_v'(t,i)^2 + \rho'(1)g_v''(t,i)\text{,}&\text{otherwise,}
\end{cases}\\
g_v''(t,i) &:= \lambda''(1)f_v'(t-1,i)^2 + \lambda'(1)f_v''(t-1,i)\\
&\quad + 2\lambda_i(i-1)f_v'(t-1,i)\text{,}\\
F_v''(t,i) &:= \begin{cases}
0\text{,}&\text{if } t = 0\\
f_v''(t,i) - \rho''(1-Q_\epsilon(t)) G_v'(t,i)^2\\
\quad - \rho'(1-Q_\epsilon(t)) G_v''(t,i)\text{,}&\text{otherwise,}
\end{cases}\\
G_v''(t,i) &:=
g_v''(t,i) -\epsilon\lambda''(P_\epsilon(t-1))F_v'(t-1,i)^2\\
&\quad - \epsilon\lambda'(P_\epsilon(t-1))F_v''(t-1,i)\\
&\quad - 2\epsilon\lambda_i(i-1)P_\epsilon(t-1)^{i-2}F_v'(t-1,i)\text{,}
\end{align*}
\begin{multline}
\mathbb{E}_t[V_i(V_i-1)P] = \epsilon L''(P_\epsilon(t)) F_v'(t,i)^2 \\
+ \epsilon L'(P_\epsilon(t)) F_v''(t,i)
+ 2 \epsilon L_i i P_\epsilon(t)^{i-1} F_v'(t,i)\text{,}\label{bv}
\end{multline}
\begin{align*}
f_c'(t,j) &:= \begin{cases}
0\text{,}&\text{if }t=0\\
\rho'(1)g_c'(t,j) + \rho_j\text{,}&\text{otherwise,}
\end{cases}\\
g_c'(t,j) &:= \lambda'(1)f_c'(t-1,j)\text{,}\\
F_c'(t,j) &:= \begin{cases}
0\text{,}&\text{if } t = 0\\
f_c'(t,j) - \rho'(1-Q_\epsilon(t))G_c'(t,j)\\
\quad - \rho_j (1-Q_\epsilon(t))^{j-1}\text{,}& \text{otherwise,}\\
\end{cases}\\
G_c'(t,j) &:=
g_c'(t,j) - \epsilon \lambda'(P_\epsilon(t-1)) F_c'(t-1,j)\text{,}
\end{align*}

\begin{align*}
f_c''(t,j) &:= \begin{cases}
0\text{,}&\text{if }t=0\\
\rho''(1)g_c'(t,j)^2 + \rho'(1)g_c''(t,j)\\
\quad + 2\rho_j(j-1)g_c'(t,j)\text{,}&\text{otherwise,}
\end{cases}\\
g_c''(t,j) &:= \lambda''(1)f_c'(t-1,j)^2 + \lambda'(1)f_c''(t-1,j)\text{,}\\
F_c''(t,j) &:= \begin{cases}
0\text{,}&\text{if } t = 0\\
f_c''(t,j) - \rho''(1-Q_\epsilon(t))G_c'(t,j)^2\\
\quad - \rho'(1-Q_\epsilon(t))G_c''(t,j)\\
\quad - 2\rho_j(j-1)(1-Q_\epsilon(t))^{j-2}\\
\quad \times G_c'(t,j)\text{,}& \text{otherwise,}\\
\end{cases}\\
G_c''(t,j) &:=
g_c''(t,j) - \epsilon \lambda''(P_\epsilon(t-1))F_c'(t-1,j)^2\\
&\quad - \epsilon \lambda'(P_\epsilon(t-1)) F_c''(t-1,j)\text{,}
\end{align*}
\begin{equation}
\mathbb{E}_t[C_j(C_j-1)P] = \epsilon L''(P_\epsilon(t)) F_c'(t,j)^2 + \epsilon L'(P_\epsilon(t)) F_c''(t,j)\text{.}\label{bc}
\end{equation}
\end{theorem}

\section{$\alpha(\epsilon,t)$ for irregular expurgated ensembles}\label{sec:gexp}
Consideration of expurgated ensembles is necessary in order to optimize performance~\cite{Am07}.
In this section, we study the irregular expurgated ensembles discussed in~\cite{1715529}, which are defined as follows: 
If there exists a single-cycle codeword, a bit chosen uniformly at random from the cycle is fixed to zero.
A $(\lambda(x),\rho(x),s)$-irregular expurgated ensemble is a $(\lambda(x),\rho(x))$-irregular ensemble
whose codewords due to single cycles of size not greater than $s$ are expurgated.
Let $\alpha(\epsilon,t,s)$ denote the coefficient of $n^{-1}$ in the bit error probability of a $(\lambda(x),\rho(x),s)$-irregular expurgated ensemble.
Accordingly, let $\beta(\epsilon,t,s)$ and $\gamma(\epsilon,t,s)$ denote the contributions to $\alpha(\epsilon,t,s)$ 
from cycle-free and single-cycle neighborhood graphs, respectively.

Due to limitation of space, we omit details 
of calculation of $\beta(\epsilon,t,s)$ and $\gamma(\epsilon,t,s)$, 
as well as the end results, and only sketch their derivation.  
For the calculation, the neighborhood graphs of depth up to $\left(t+\left\lfloor\frac{s+1}{2}\right\rfloor\right)$ should be considered
since cycles which are included in neighborhood graphs of depth $\left(t+\left\lfloor\frac{s+1}{2}\right\rfloor\right)$ will affect 
the bit error probability $\Pb(n,\epsilon,t)$ via expurgation.  
The coefficients $\beta(\epsilon,t,s)$ and $\gamma(\epsilon,t,s)$ are 
calculated recursively, in ways similar to the calculations 
of $\beta(\epsilon,t)$ and $\gamma(\epsilon,t)$, respectively.

\section{Numerical calculations and simulations}\label{calsim}
\begin{figure}[ht]
\psfrag{e}{$\epsilon$}
\psfrag{alpha}{$\alpha(\epsilon,t)$}
\includegraphics[width=\hsize]{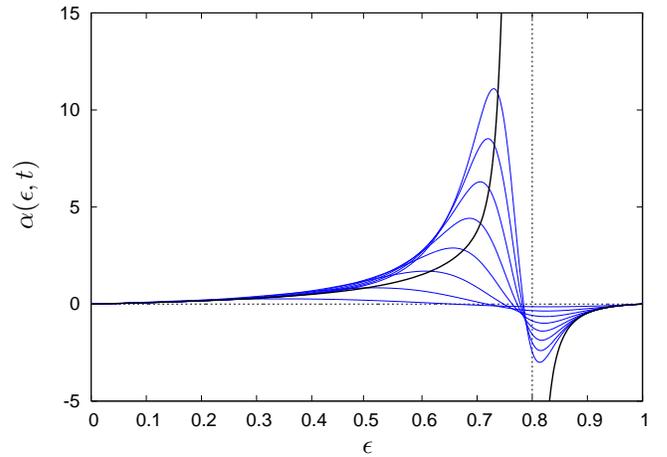}
\caption{Calculation results for an irregular unexpurgated ensemble.
$\lambda(x) = 0.500 x + 0.153 x^2 + 0.112 x^3 + 0.055 x^4 + 0.180 x^8$,
$\rho(x) = 0.492 x^2 + 0.508 x^3$.
Thin curves show results for $t=1,2,\dotsc,8$. Thick curve shows result for $t=50$. The threshold is about $0.8$.
}
\label{irgalpha}
\end{figure}

\begin{figure}[ht]
\psfrag{e}{$\epsilon$}
\psfrag{y}{$n\left|\Pb(n,\epsilon,t)-\Pb(\infty,\epsilon,t)\right|$}
\psfrag{alpha}{\footnotesize \hspace{-2em}$\alpha(\epsilon,t)$}
\includegraphics[width=\hsize]{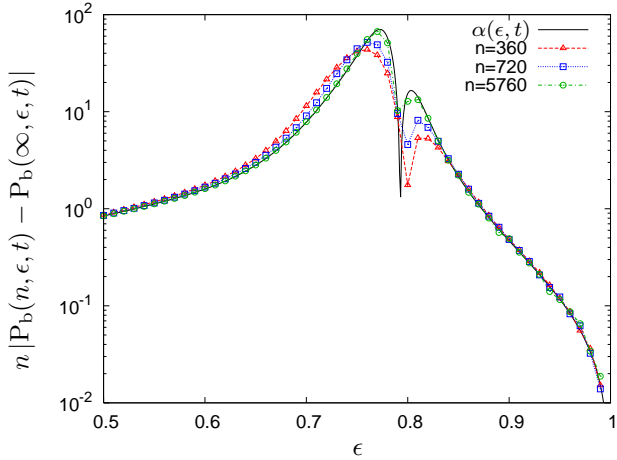}
\caption{Simulation results for an irregular unexpurgated ensemble.
$\lambda(x) = 0.500 x + 0.153 x^2 + 0.112 x^3 + 0.055 x^4 + 0.180 x^8$,
$\rho(x) = 0.492 x^2 + 0.508 x^3$.
Blocklengths are  $360$, $720$ and $5760$.
The number of iterations is $20$.
}
\label{irgsim}
\end{figure}

\begin{figure}[th]
\psfrag{x}{$\epsilon$}
\psfrag{y}{$\alpha(\epsilon,t,s)$}
\includegraphics[width=\hsize]{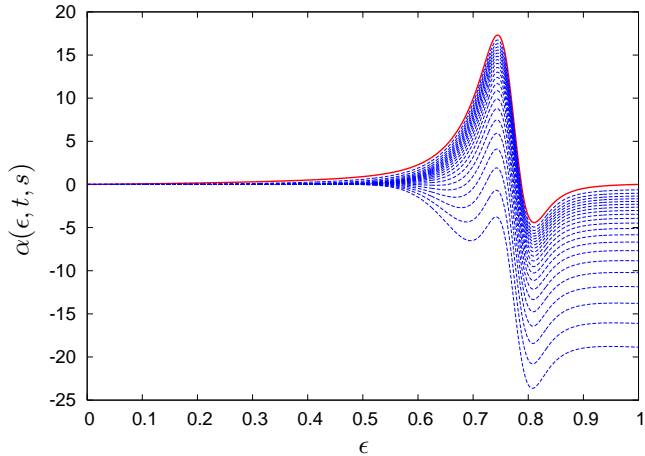}
\caption{Calculation results for irregular expurgated ensembles.
$\lambda(x) = 0.500 x + 0.153 x^2 + 0.112 x^3 + 0.055 x^4 + 0.180 x^8$,
$\rho(x) = 0.492 x^2 + 0.508 x^3$.
The number of iterations is $10$.
The top curve shows results for an irregular unexpurgated ensemble.
The other curves are results for irregular expurgated ensembles where the size of expurgations varies from $1$ to $21$.
}
\label{expalpha}
\end{figure}

In this section, calculation results of $\alpha(\epsilon,t)$ for irregular unexpurgated ensembles, and of $\alpha(\epsilon,t,s)$ for irregular expurgated ensembles,
and simulation results for irregular unexpurgated ensembles are shown.
Figure~\ref{irgalpha} shows calculation results of $\alpha(\epsilon,t)$ for an optimized (via density evolution~\cite{258573}) irregular unexpurgated ensemble.
It seems to converge for all $\epsilon$ except the threshold.
Although the limit $\lim_{t\to\infty}\alpha(\epsilon,t)$ is obtained for regular ensembles~\cite{4658691},
it has not been obtained for irregular ensembles, nor a proof of the convergence.
Figure~\ref{irgsim} shows $\left|\alpha(\epsilon,t)\right|$ and
simulation results of $\left|n(\Pb(n,\epsilon,t)-\Pb(\infty,\epsilon,t))\right|$ which should converge to $\left|\alpha(\epsilon,t)\right|$
as $n\to\infty$.
The simulation results are almost the same as the limit $\alpha(\epsilon,t)$.
It is practically plausible but theoretically very strange, 
since $\alpha(\epsilon,t)$ consists of contributions of cycle-free and single-cycle neighborhood graphs,
whereas the probabilities of cycle-free and single-cycle neighborhood graphs are effectively zero when the blocklength is $5760$ and the number of iterations is $20$.
Figure~\ref{expalpha} shows calculation results of $\alpha(\epsilon,t,s)$ for an irregular expurgated ensemble.
The coefficient $\alpha(\epsilon,t,s)$ of $n^{-1}$ decreases as expurgation size $s$ increases.
Unfortunately, we can not simulate irregular expurgated ensembles due to its high computational costs.

\section{Conclusion and future works}\label{cncld}
The coefficient $\alpha(\epsilon,t)$ of $n^{-1}$ in the bit error probability for irregular expurgated ensembles are obtained.
A reason of the strangely fast convergence to $\alpha(\epsilon,t)$ is an open problem.
Finite-length and finite-iteration optimization is an important future work.
Furthermore, generalization to binary memoryless symmetric channels remains to be done. 


\enlargethispage{-25em}

\section*{Acknowledgment}
TT acknowledges support of the Grant-in-Aid for Scientific Research 
on Priority Areas (No.~18079010), MEXT, Japan.

\bibliographystyle{IEEEtran}
\bibliography{IEEEabrv,ldpc}


\end{document}